\documentclass[11pt]{article}
\usepackage{fullpage}
\usepackage{amsmath}
\usepackage{amssymb}
\usepackage{amsfonts}
\usepackage{epsfig}
\usepackage{epic}
\usepackage{curves}
\usepackage[dvips]{color}

\newcounter{algoctr}

\newif\ifnotesw\noteswtrue
   {\ifnotesw\marginpar[\hfill\(\top\)]{\(\top\)}\fi}%
   {\ifnotesw\marginpar[\hfill\(\bot\)]{\(\bot\)}\fi}

\newcommand{\mnote}[1]%
    {\ifnotesw\marginpar%
        [{\scriptsize\begin{minipage}[t]{\marginparwidth}
        \raggedleft#1%
                        \end{minipage}}]%
        {\scriptsize\begin{minipage}[t]{\marginparwidth}
        \raggedright#1%
                        \end{minipage}}%
    \fi}

\newcommand{\ignore}[1]{}

\newcommand{\etal}{{\it et al.~}}

\newsavebox{\given}
\savebox{\given}[1em]{\rule[-1.5ex]{.2mm}{4ex}}

\newtheorem{theorem}{Theorem}
\newtheorem{corollary}[theorem]{Corollary}

\newtheorem{fact}[theorem]{Fact}

\newcommand{\blackslug}{\rule{7pt}{7pt}}

\newcommand{\iverson}[1]{\lbrack\!\lbrack #1 \rbrack\!\rbrack}

\newcommand{\qed}{\hfill{\setlength{\fboxsep}{0pt}
\framebox[7pt]{\rule{0pt}{7pt}}}}

\renewcommand{\notin}{\ifmmode \not\in \else $\not\in$ \fi}
\newlength{\thislabel}
\newcommand{\labsize}[1]{\settowidth{\thislabel}{#1}}

\newcommand{\prf}{\par\noindent{\sl Proof } \hspace{.01 in}}
\newcommand{\zo}{\{0,1\}}

\def\dotdef{\stackrel{.}{=}}

\newcommand{\bra}[1]{\langle #1 |}
\newcommand{\ket}[1]{| #1 \rangle}
\newcommand{\braket}[2]{\langle #1 | #2 \rangle}

\newcommand{\tstar}{t^{\star}}

\newcommand{\idx}{\xi}

\title{
On quantum perfect state transfer in weighted join graphs
} 
\author{
{Ricardo Javier Angeles-Canul}\footnote{Centro de Investigaci\'{o}n en Matem\'{a}ticas, Universidad Aut\'{o}noma del Estado de Hidalgo, Pachuca, Hidalgo, Mexico. email: richywhitedragon@gmail.com}
\and
{Rachael M. Norton}\footnote{Dept. Mathematics, Bowdoin College, Brunswick, Maine, U.S.A. email: rnorton@bowdoin.edu}
\and
{Michael C. Opperman}\footnote{Dept. Mathematics and Computer Science, Clarkson University, Potsdam, New York, U.S.A. email: oppermmc@clarkson.edu}
\and
{Christopher C. Paribello}\footnote{Dept. Mathematics and Computer Science, Clarkson University, Potsdam, New York, U.S.A. email: paribecc@clarkson.edu}
\and
{Matthew C. Russell}\footnote{Dept. Mathematics, Taylor University, Upland, Indiana, U.S.A. email: matthew\_russell@taylor.edu}
\and
{Christino Tamon}\footnote{Dept. Mathematics and Computer Science, Clarkson University, Potsdam, New York, U.S.A. email: tino@clarkson.edu. Corresponding author.}
}

\date{\today}

\begin{document}
\bibliographystyle{plain}
\maketitle

\begin{abstract}
We study perfect state transfer on quantum networks represented by weighted graphs.
Our focus is on graphs constructed from the join and related graph operators. Some specific 
results we prove include:
\begin{itemize}
\item The join of a weighted two-vertex graph with any regular graph has perfect state transfer.
	This generalizes a result of Casaccino \etal \cite{clms09} where the regular graph 
	is a complete graph or a complete graph with a missing link. In contrast, the half-join of 
	a weighted two-vertex graph with any weighted regular graph has no perfect state transfer. 
	This implies that adding weights in a complete bipartite graph do not help in achieving
	perfect state transfer.
\item A Hamming graph has perfect state transfer between each pair of its vertices. This is obtained
	using a closure property on weighted Cartesian products of perfect state transfer graphs. Moreover, 
	on the hypercube, we show that perfect state transfer occurs between uniform superpositions on
	pairs of arbitrary subcubes. This generalizes results of Bernasconi \etal \cite{bgs08} and 
	Moore and Russell \cite{mr02}.
\end{itemize}
Our techniques rely heavily on the spectral properties of graphs built using the join and Cartesian product
operators.
\vspace{.045in}
\par\noindent{\em Keywords}: Perfect state transfer, quantum networks, weighted graphs, join.
\end{abstract}



\section{Introduction}

Recently, the notion of perfect state transfer in quantum networks modeled by graphs has received 
considerable attention in quantum information \cite{cdel04,cddekl05,sss07,bgs08,bcms09,bp09,clms09}. 
A main goal in this line of research is to find and characterize graph structures which exhibit 
perfect state transfer between pairs of vertices in the graph. 
This is a useful property of quantum networks since it facilitates information transfer between locations.

We may conveniently view the perfect state transfer problem in the context of quantum walks on graphs 
\cite{fg98,k06}. In this setting, the initial state of the quantum system is described by a unit vector 
on some initial vertex $a$. To achieve perfect transfer to a target vertex $b$ at time $t$, the quantum walk 
amplitude of the system at time $t$ on vertex $b$ must be of unit magnitude. In other words, we require that
$|\bra{b}e^{-itA_{G}}\ket{a}| = 1$,
where $A_{G}$ is the adjacency matrix of the underlying graph $G$ that describes the quantum network. 

Christandl \etal \cite{cdel04} observed that the Cartesian products of paths of length three (two-link hypercubes)
admit perfect state transfer between antipodal vertices. They also noted that paths of length four or larger 
do not possess perfect state transfer unless their edges are weighted in a specific manner (see \cite{cddekl05}). 
In fact, this weighting scheme corresponds closely to the hypercube structure.
This crucially shows that edge weights can be useful in achieving perfect state transfer 
on graphs which are known not to possess the property.

It is known that complete graphs do not have perfect state transfer.
But surprisingly, Casaccino \etal \cite{clms09} observed that adding weighted self-loops on 
two vertices in a complete graph helps create perfect state transfer between the two vertices. 
We generalize their observation by considering the join of a weighted two-vertex graph with an 
arbitrary regular graph. 
We prove that adding weights also helps for perfect state transfer in this more general case. 
On the other hand, we show that the half-join between a weighted two-vertex graph with a weighted self-join of 
an arbitrary  regular graph, where each vertex of the two-vertex graph is connected to exactly half of the join
graph, has no perfect state transfer for any set of weights. This implies that weights provably do not help
in achieving perfect state transfer in a complete bipartite graphs.
The full connection that is available in the standard join seems crucial in achieving perfect state transfer.

Bernasconi \etal \cite{bgs08} gave a complete characterization of perfect state transfer on the hypercubes. 
They proved that perfect state transfer is possible at time $t = \pi/2$ between any pair of vertices. 
We will refer to this stronger property as universal perfect state transfer.
Previously known results on perfect state transfer on other graphs, such as integral circulants \cite{bp09} 
and two-link hypercubes \cite{cdel04}, only allow perfect state transfer between antipodal 
vertices (which are vertices at maximum distance from each other). 
Recent results on integral circulants and other graphs (see \cite{anoprt09}) have exhibited
perfect state transfer between non-antipodal vertices, but most of these graphs still lack 
the universal perfect state transfer property.

We show that weights are useful for universal perfect state transfer in the family of Hamming graphs, 
which is a generalization of the hypercube family. We prove this result by extending 
the observation of Christandl \etal \cite{cdel04} to perfect state transfer on weighted Cartesian products. 
For a weighted $n$-cube, we prove a stronger universal perfect state transfer property. We show that 
perfect state transfer occurs between uniform superpositions over two arbitrary subcubes of the $n$-cube. 
This generalizes the results of both Bernasconi \etal \cite{bgs08} mentioned above and also of 
Moore and Russell \cite{mr02} on the uniform mixing of a quantum walk on the $n$-cube. 
We note that Bernasconi \etal \cite{bgs08} proved universal perfect state transfer on the $n$-cube by 
{\em dynamically} changing the underlying hypercubic structure of the graph. In contrast, our scheme is 
based on {\em static} weights which can be interpreted dynamically with time.

Note that if we allow zero edge weights then universal perfect state transfer becomes trivial.
Assuming that the two source and target vertices are connected, find a path connecting them, assign the
hypercubic weights to the edges on this path (as in Christandl \etal \cite{cdel04}) and zero weights to the other
edges. This shows that universal perfect state transfer can be achieved if zero edge weights are allowed.

Our work exploits the machinery developed in \cite{anoprt09} and their extensions to weighted graphs. 
These include the join theorem for regular graphs and the closure property for Cartesian product of 
perfect state transfer graphs.


\section{Preliminaries}

For a logical statement $\mathcal{S}$, the Iversonian notation $\iverson{\mathcal{S}}$ is $1$ 
if $\mathcal{S}$ is true and $0$ otherwise (see Graham, Knuth and Patashnik \cite{gkp94}).
As is standard, we use $I_{n}$ and $J_{n}$ to denote the $n \times n$ identity and all-one matrices, 
respectively; we drop the subscript $n$ whenever the context is clear. 

The graphs $G=(V,E)$ we study are finite, mostly simple, undirected, and connected.
The adjacency matrix $A_{G}$ of a graph $G$ is defined as $A_{G}[u,v] = \iverson{(u,v) \in E}$.
A graph $G$ is called $k$-regular if each vertex has $k$ adjacent neighbors. That is,
the neighbor set $\{v \in V : (u,v) \in E\}$ of $u$ has cardinality $k$ for each vertex $u \in V$.
In most cases, we also require $G$ to be vertex-transitive, that is, for any $a,b \in V$, there is an 
automorphism $\pi \in Aut(G)$ with $\pi(a)=b$. 

In this paper, we also consider edge-weighted graphs $\widetilde{G}=(V,E,w)$, where 
$w: E \rightarrow \mathbb{R}$ is a function that assigns weights to edges. 
In the simplest case, we take an unweighted graph $G=(V,E)$ and add self-loops with weight
$\alpha$ to all vertices and assign a weight of $\beta$ to all edges; we denote such a graph
by $\widetilde{G}(\alpha,\beta)$. Note that the adjacency matrix of $\widetilde{G}$ is given by
$\alpha I + \beta A_{G}$. Unless otherwise stated, most of our weighted graphs will be of this form.

We denote the complete graph on $n$ vertices by $K_{n}$.
The Cartesian product $G \oplus H$ of graphs $G$ and $H$ is a graph whose adjacency matrix is
$I \otimes A_{H} + A_{G} \otimes I$ (see Lov\'{a}sz \cite{lovasz}, page 617).
The binary $n$-dimensional hypercube $Q_{n}$ may be defined recursively as 
$Q_{n} = K_{2} \oplus Q_{n-1}$, for $n \ge 2$, and $Q_{1} = K_{2}$.
Similarly, the Hamming graph $H(q,n)$ is defined as $K_{q}^{\oplus n}$; this may be viewed as
a $q$-ary $n$-dimensional hypercube.

The {\em join} $G + H$ of graphs $G$ and $H$ is defined as $\overline{G+H} = \overline{G} \cup \overline{H}$; 
that is, we take a copy of $G$ and a copy of $H$ and connect all vertices of $G$ with all vertices of $H$
(see \cite{sw78}). We will also consider the weighted join $G +_{\rho} H$ where we assign a weight of $\rho$
to the edges that connect $G$ and $H$; more specifically, the adjacency matrix of $G +_{\rho} H$ is given by 
\begin{equation}
\begin{bmatrix} A_{G} & \rho J \\ \rho J & A_{H} \end{bmatrix},
\end{equation}
with the appropriate dimensions on the two all-one $J$ matrices.
A {\em cone} on a graph $G$ is the graph $K_{1} + G$. Similarly, a connected {\em double} cone on a graph $G$ 
is the graph $K_{2} + G$; similarly, a disconnected double cone is the graph $\overline{K}_{2} + G$. 
When $G$ is the empty graph, the connected double-cone is simply the complete graph whereas the disconnected
double-cone is the complete graph with a missing edge (see \cite{bcms09,clms09}).
On the other hand, a connected (or disconnected) double {\em half-cone} on a graph $G$ is 
formed by taking $K_{2}$ (or $\overline{K}_{2}$) and $G + G$ and connecting each vertex of the two-vertex graph to 
exactly one copy of $G$ in the join $G+G$. 
When $G$ is the empty graph, the double half-cone simply yields a complete bipartite graph.
For more background on algebraic graph theory, we refer the reader to the monograph by Biggs \cite{biggs}.

For a graph $G=(V,E)$, let $\ket{\psi(t)} \in \mathbb{C}^{|V|}$ be a time-dependent amplitude vector 
over $V$. The continuous-time quantum walk on $G$ is defined using Schr\"{o}dinger's equation as
\begin{equation}
\ket{\psi(t)} = e^{-it A_{G}} \ket{\psi(0)},
\end{equation}
where $\ket{\psi(0)}$ is the initial amplitude vector (see \cite{fg98}). 
Further background on quantum walks on graphs can be found in the survey by Kendon \cite{k06}.
We say $G$ has {\em perfect state transfer} (PST) from vertex $a$ to vertex $b$ at time $\tstar$ if
\begin{equation} \label{eqn:pst}
|\bra{b}e^{-i\tstar A_{G}}\ket{a}| = 1,
\end{equation}
where $\ket{a}$, $\ket{b}$ denote the unit vectors corresponding to the vertices $a$ and $b$,
respectively. The graph $G$ has perfect state transfer if there exist distinct vertices $a$ and $b$ 
in $G$ and a time $\tstar \in \mathbb{R}^{+}$ so that (\ref{eqn:pst}) is true. We say that $G$ has
{\em universal} perfect state transfer if (\ref{eqn:pst}) occurs between all distinct pairs of
vertices $a$ and $b$ of $G$.

\subsection{Example: Triangle}

We begin by describing an explicit example of the role of weights for perfect state transfer in a
triangle, or $K_{3}$, which is the complete graph on three vertices.
The eigenvalues of $K_{3}$ are $2$ (simple) and $-1$ (with multiplicity two) with eigenvectors
$\ket{F_{k}}$, where $\ket{F_{k}}$ are the columns of the Fourier matrix, 
with $\braket{j}{F_{k}} = \omega_{3}^{jk}/\sqrt{3}$, for $j,k \in \{0,1,2\}$ (see Biggs \cite{biggs}).
The quantum walk on $K_{3}$ yields
\begin{equation}
\bra{1}e^{-itK_{3}}\ket{0} 
	= \bra{1}\left\{\sum_{k=0}^{2} e^{-it\lambda_{k}}\ket{F_{k}}\bra{F_{k}}\right\}\ket{0} 
	= -\frac{2}{3}ie^{-it/2}\sin(3t/2).
\end{equation}

\begin{figure}[t]
\begin{center}
\setlength{\unitlength}{0.75cm}
\begin{picture}(15,5)
\thicklines
\put(2,1){\circle*{0.35}}
\put(2,3){\circle*{0.35}}
\put(4,2){\circle*{0.35}}
\put(2,1){\color{blue}{\line(0,1){2.0}}}
\put(2,1){\line(2,1){2.0}}
\put(2,3){\line(2,-1){2.0}}
\put(1.65,1){\color{red}{\bigcircle{0.75}}}
\put(1.65,3){\color{red}{\bigcircle{0.75}}}
\put(0.75,0.9){\text{$\mu$}}
\put(0.75,2.9){\text{$\mu$}}
\put(1.5,1.9){\text{$\eta$}}

\put(9,1){\circle*{0.35}}
\put(9,3){\circle*{0.35}}
\put(11,1){\circle*{0.35}}
\put(11,3){\circle*{0.35}}
\put(13,1){\circle*{0.35}}
\put(13,3){\circle*{0.35}}
\put(9,1){\color{blue}{\line(0,1){2.0}}}
\put(9,1){\line(1,0){2.0}}
\put(9,1){\line(2,1){4.0}}
\put(9,1){\line(1,1){2.0}}
\put(9,3){\line(1,0){4.0}}
\put(9,3){\line(2,-1){4.0}}
\put(9,3){\line(1,-1){2.0}}
\put(8.65,1){\color{red}{\bigcircle{0.75}}}
\put(8.65,3){\color{red}{\bigcircle{0.75}}}
\put(7.75,0.9){\text{$\mu$}}
\put(7.75,2.9){\text{$\mu$}}
\put(8.5,1.9){\text{$\eta$}}
\curve(9,1,11,0.5,13,1)
\curve(9,3,11,3.5,13,3)
\put(11,1){\line(0,1){2.0}}
\put(11,1){\line(1,0){2.0}}
\put(13,1){\line(0,1){2.0}}
\put(11,3){\line(1,0){2.0}}
\end{picture}
\caption{Weighted joins: (a) $K_{2} + K_{1}$ (b) $K_{2} + C_{4}$.
Perfect state transfer occurs between the weighted self-loop vertices. 
Without the self-loops and weights, there is no perfect state transfer (see \cite{clms09}).}
\end{center}
\end{figure}
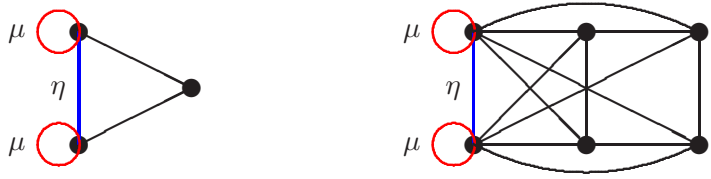

So, it is clear that there is no perfect state transfer on $K_{3}$ (see \cite{abtw03,clms09}).
Now, consider adding self-loops on vertices $0$ and $1$ with weight $\mu$ and putting a weight of
$\eta$ on the edge connecting $0$ and $1$. The adjacency matrix of this weighted $\widetilde{K}_{3}$ is
\begin{equation}
\widetilde{K}_{3} = 
	\begin{bmatrix} 
	\mu & \eta & 1 \\ 
	\eta & \mu & 1 \\ 
	1 & 1 & 0 
	\end{bmatrix}
\end{equation}
The spectra of $\widetilde{K}_{3}$ is given by the eigenvalues
$\lambda_{0}=\mu-\eta$ and $\lambda_{\pm} = 2\alpha_{\pm}$, where $\alpha_{\pm} = \frac{1}{4}(\delta \pm \Delta)$, 
$\delta = \mu+\eta$ and $\Delta = \sqrt{\delta^{2}+8}$, with corresponding orthonormal eigenvectors
\begin{equation}
\ket{v_{0}} = \frac{1}{\sqrt{2}}\begin{bmatrix} 1 \\ -1 \\ 0 \end{bmatrix}, \ \ \
\ket{v_{\pm}} = \frac{1}{\sqrt{2\alpha^{2}_{\pm}+1}}\begin{bmatrix} \alpha_{\pm} \\ \alpha_{\pm} \\ 1 \end{bmatrix}
\end{equation}
The perfect state transfer equation between the two vertices with weighted self-loops are given by
\begin{eqnarray}
\bra{1}e^{-it\widetilde{K}_{3}}\ket{0}
	& = & \bra{1}e^{-it\widetilde{K}_{2}}\ket{0}
		+ \frac{1}{2}e^{-it\delta}\left\{ e^{it\delta/2}
			\left[\cos\left(\frac{\Delta}{2}t\right) - 
				i\frac{\delta}{\Delta}\sin\left(\frac{\Delta}{2}t\right)\right] - 1 \right\},
\end{eqnarray}
where $\widetilde{K}_{2}$ is 
$\widetilde{K}_{2}(\mu,\eta)$.
Recall that the perfect state transfer $\bra{1}e^{-itK_{2}}\ket{0}$ on 
the (unweighted) $K_{2}$ is given by $-i\sin(t)$. Thus, the weighted $\widetilde{K}_{2}$ 
has perfect state transfer at time $\tstar = (2\mathbb{Z}+1)\pi/2\eta$, since the self-loop weight $\mu$
disappears into an irrelevant phase factor and the edge weight $\eta$ translates into a time-scaling.
So, to achieve perfect state transfer on $\widetilde{K}_{3}$, it suffices to have
\begin{equation}
\cos\left( \frac{\delta}{4\eta} \pi \right)\cos\left( \frac{\Delta}{4\eta} \pi \right) = 1.
\end{equation}
Equivalently, we require that:
\begin{enumerate}
\item $A \dotdef \delta/4\eta$ be an integer;
\item $B \dotdef \Delta/4\eta$ be an integer; and
\item $A \equiv B\pmod{2}$ or that $A$ and $B$ have the same parity.
\end{enumerate}
From the first two conditions, we require that $\delta/\Delta$ be a rational number
$p/q < 1$ with $\gcd(p,q)=1$. 
Restating this last condition on $p$ and $q$ and simplifying, we get that
\begin{equation}
\delta \ = \ p \ \sqrt{\frac{8}{q^{2}-p^{2}}}, \hspace{.5in}
\Delta \ = \ q \ \sqrt{\frac{8}{q^{2}-p^{2}}}.
\end{equation}
So, we may choose
\begin{equation}
\eta \ = \ \frac{1}{4}\sqrt{\frac{8}{q^{2}-p^{2}}}
\end{equation}
so that both $\delta/4\eta$ and $\Delta/4\eta$ are integers.
Therefore, we choose odd integers $p$ and $q$ satisfying $\gcd(p,q)=1$;
this will satisfy all three conditions stated above.
This shows that there are infinitely many weights $\mu$ and $\eta$ (via infinitely choices of odd integers $p$ and $q$) 
which allow perfect state transfer on $\widetilde{K}_{3}$.
We generalize this example in our join theorem for arbitrary regular weighted graphs.

This example complements a result of Casaccino \etal \cite{clms09} which showed 
the power of weighted self-loops on complete graphs. Our analysis above shows that perfect state
transfer is achieved through edge weights instead.

\section{Join of weighted regular graphs}

In this section, we prove that the existence of perfect state transfer in a join of two arbitrary 
regular weighted graphs can be reduced to perfect state transfer in one of the graphs. In fact, since
we add weights to our graphs in a particular way, this is a reduction onto the unweighted version of one 
of the graphs. This allows us to analyze the double-cone on any regular graph; that is, the join of 
$K_{2}$ with an arbitrary regular graph. 
The next theorem is a generalization of a similar join theorem given in \cite{anoprt09}.

\begin{theorem} \label{thm:weighted-binary-join}
For $j \in \{1,2\}$, let $\widetilde{G}_{j}(\mu_{j},\eta_{j})$ be a $k_{j}$-regular graph 
on $n_{j}$ vertices, where each vertex has a self-loop with weight $\mu_{j}$ and each edge 
has weight $\eta_{j}$. Also, for $j \in \{1,2\}$, let 
\begin{equation}
\kappa_{j} = \mu_{j}+\eta_{j}k_{j}.
\end{equation}
Suppose that $a$ and $b$ are two vertices in $\widetilde{G}_{1}$.
Let $\mathcal{G} = \widetilde{G}_{1}(\mu_{1},\eta_{1}) + \widetilde{G}_{2}(\mu_{2},\eta_{2})$ 
be the join of the weighted graphs. Then, 
\begin{equation} \label{eqn:pst-join-reduction}
\bra{b}e^{-it A_{\mathcal{G}}}\ket{a}
	= \bra{b} e^{-it A_{\widetilde{G}_{1}}} \ket{a} 
	+
	\frac{e^{-it \kappa_{1}}}{n_{1}}
	\left\{e^{it\delta/2} \left[ \cos\left(\frac{\Delta t}{2}\right) 
	- i\left(\frac{\delta}{\Delta}\right)\sin\left(\frac{\Delta t}{2}\right) \right] - 1 \right\}
\end{equation}
where 
$\delta = \kappa_{1}-\kappa_{2}$ and $\Delta = \sqrt{\delta^{2} + 4n_{1}n_{2}}$.
\end{theorem}
\prf
Let $G_{j}$ be the simple and unweighted version of $\widetilde{G}_{j}$, for $j \in \{1,2\}$; 
that is, $G_{j} = \widetilde{G}_{j}(0,1)$. 
Whenever it is clear from context, we denote $\widetilde{G}_{j}(\mu_{j},\eta_{j})$ as simply 
$\widetilde{G}_{j}$.

If $\lambda_{k}$ and $\ket{u_{k}}$ are the eigenvalues and eigenvectors of $A_{G_{1}}$, for $k=0,\ldots,n_{1}-1$,
then
\begin{equation}
\bra{b}e^{-it A_{G_{1}}}\ket{a} = 
	\bra{b}\left\{\sum_{k=0}^{n_{1}-1} \ket{u_{k}}\bra{u_{k}} e^{-it\lambda_{k}}\right\} \ket{a}.
\end{equation}
Here, we assume $\ket{u_{0}}$ is the all-one eigenvector (that is orthogonal to the other eigenvectors) 
with eigenvalue $\lambda_{0} = k_{1}$. 
By the same token, let $\theta_{\ell}$ and $\ket{v_{\ell}}$ be the eigenvalues and eigenvectors of $A_{G_{2}}$, 
for $\ell = 0,\ldots,n_{2}-1$. Also, let $\ket{v_{0}}$ be the all-one eigenvector 
(with eigenvalue $\theta_{0} = k_{2}$) which is orthogonal to the other eigenvectors $\ket{v_{\ell}}$, $\ell \neq 0$.

Let $\mathcal{G} = \widetilde{G}_{1} + \widetilde{G}_{2}$. Note that the adjacency matrix of $\mathcal{G}$ is
\begin{equation}
A_{\mathcal{G}} = 
	\begin{bmatrix} 
	\mu_{1}I+\eta_{1}A_{G_{1}} & J_{n_{1} \times n_{2}} \\ 
	J_{n_{2} \times n_{1}} & \mu_{2}I+\eta_{2}A_{G_{2}} 
	\end{bmatrix}.
\end{equation}
Let $\delta = \kappa_{1}-\kappa_{2}$, where $\kappa_{j} = \mu_{j}+\eta_{j}k_{j}$, for $j \in \{1,2\}$.
The eigenvalues and eigenvectors of $A_{\mathcal{G}}$ are given by the following three sets:
\begin{itemize}
\item 
	For $k=1,\ldots,n_{1}-1$, let $\ket{u_{k},0_{n_{2}}}$ be a column vector formed by concatenating 
	the column vector $\ket{u_{k}}$ with the zero vector of length $n_{2}$. 
	Then, $\ket{u_{k},0_{n_{2}}}$ is an eigenvector with eigenvalue 
	$\widetilde{\lambda}_{k} = \mu_{1}+\eta_{1}\lambda_{k}$.
	Note that $\widetilde{\lambda}_{0} = \kappa_{1}$.

\item
	For $\ell=1,\ldots,n_{2}-1$, let $\ket{0_{n_{1}},v_{\ell}}$ be a column vector formed by concatenating 
	the zero vector of length $n_{1}$ with the column vector $\ket{v_{\ell}}$.
	Then, $\ket{0_{n_{1}},v_{\ell}}$ is an eigenvector with eigenvalue 
	$\widetilde{\theta}_{\ell} = \mu_{2}+\eta_{2}\theta_{\ell}$.

\item
	Let $\ket{\pm} = \frac{1}{\sqrt{L_{\pm}}}\ket{\alpha_{\pm},1_{n_{2}}}$ be a column vector formed
	by concatenating the vector $\alpha_{\pm}\ket{1_{n_{1}}}$ with the vector $\ket{1_{n_{2}}}$, where
	$\ket{1_{n_{1}}}$, $\ket{1_{n_{2}}}$ denote the all-one vectors of length $n_{1}$, $n_{2}$, respectively.
	Then, $\ket{\pm}$ is an eigenvector with eigenvalue 
	$\widetilde{\lambda}_{\pm} = n_{1}\alpha_{\pm} + \kappa_{2}$.
	Here, we have
	\begin{equation}
	\alpha_{\pm} = \frac{1}{2n_{1}}(\delta \pm \Delta), \ \ \ 
	\Delta^{2} = \delta^{2} + 4n_{1}n_{2}, \ \ \
	L_{\pm} = n_{1}(\alpha_{\pm})^{2} + n_{2},
	\end{equation}

\end{itemize}
In what follows, we will abuse notation by using $\ket{a}$, $\ket{b}$ for both 
$\widetilde{G}_{1}$ and $\widetilde{G}_{1}+\widetilde{G}_{2}$;
their dimensions differ in both cases, although it will be clear from context which version is used.
The quantum wave amplitude from $a$ to $b$ is given by
\begin{eqnarray} 
\bra{b}e^{-it A_{\mathcal{G}}}\ket{a}
	& = & \bra{b} e^{-it A_{\mathcal{G}}} \left\{\sum_{k=1}^{n_{1}-1}\braket{u_{k},0_{n_{2}}}{a} \ket{u_{k},0_{n_{2}}} 
			+ 
			\sum_{\pm} \frac{\alpha_{\pm}}{\sqrt{L_{\pm}}} \ket{\pm} \right\} \\
	& = & \bra{b} \left\{\sum_{k=1}^{n_{1}-1}\braket{u_{k}}{a} e^{-it\widetilde{\lambda}_{k}}\ket{u_{k},0_{n_{2}}} + 
			\sum_{\pm} \frac{\alpha_{\pm}}{\sqrt{L_{\pm}}} e^{-it\widetilde{\lambda}_{\pm}} \ket{\pm} \right\} \\
	& = & \sum_{k=1}^{n_{1}-1}\braket{b}{u_{k}}\braket{u_{k}}{a} e^{-it\widetilde{\lambda}_{k}} + 
			\sum_{\pm} \frac{\alpha_{\pm}^{2}}{L_{\pm}} e^{-it\widetilde{\lambda}_{\pm}}.
\end{eqnarray}
This shows that
\begin{eqnarray}
\bra{b}e^{-it A_{\mathcal{G}}}\ket{a}
\label{eqn:pst-join}
	& = & \bra{b}\left\{\sum_{k=0}^{n_{1}-1}\ket{u_{k}}\bra{u_{k}} e^{-it\widetilde{\lambda}_{k}}\right\} \ket{a} 
			-\frac{e^{-it \kappa_{1}}}{n_{1}} + 
			\sum_{\pm} \frac{\alpha_{\pm}^{2}}{L_{\pm}} e^{-it\widetilde{\lambda}_{\pm}} \\
\label{eqn:pst-join2}
	& = & \bra{b} e^{-it A_{\widetilde{G}_{1}}} \ket{a} 
			+ \sum_{\pm} \frac{\alpha_{\pm}^{2}}{L_{\pm}} e^{-it\widetilde{\lambda}_{\pm}} 
			-\frac{e^{-it \kappa_{1}}}{n_{1}}.
\end{eqnarray}
To analyze the second term next, we use the following identities whose correctness follows easily from the
definitions of $\alpha_{\pm}$, $L_{\pm}$, $\delta$ and $\Delta$:
\begin{eqnarray}
\alpha_{+}\alpha_{-} & = & -(n_{2}/n_{1}) \\
\alpha_{+} + \alpha_{-} & = & \delta/n_{1} \\
L_{+}L_{-} & = & (n_{2}/n_{1})\Delta^{2} \\
L_{+} + L_{-} & = & \Delta^{2}/n_{1} \\
(\alpha_{\pm})^{2}L_{\mp} & = & (n_{2}/n_{1})L_{\pm} \\
\label{eqn:lambda-pm}
\widetilde{\lambda}_{\pm} & = & (\hat{\delta} \pm \Delta)/2
\end{eqnarray}
where $\hat{\delta} = \kappa_{1} + \kappa_{2}$. 
Therefore, the summand in (\ref{eqn:pst-join2}) is given by
\begin{eqnarray}
\sum_{\pm} \frac{\alpha_{\pm}^{2}}{L_{\pm}} e^{-it\widetilde{\lambda}_{\pm}} 
	& = & \frac{1}{n_{1}} e^{-it\hat{\delta}/2} \left[ \cos\left(\frac{\Delta t}{2}\right) 
				- i\left(\frac{\delta}{\Delta}\right)\sin\left(\frac{\Delta t}{2}\right) \right].
\end{eqnarray}
This yields
\begin{equation}
\bra{b}e^{-it A_{\mathcal{G}}}\ket{a}
	= \bra{b} e^{-it A_{\widetilde{G}_{1}}} \ket{a} 
	+
	\frac{e^{-it \kappa_{1}}}{n_{1}}
	\left\{e^{it\delta/2} \left[ \cos\left(\frac{\Delta t}{2}\right) 
	- i\left(\frac{\delta}{\Delta}\right)\sin\left(\frac{\Delta t}{2}\right) \right] - 1 \right\}
\end{equation}
which proves the claim.
\qed\\

\par\noindent We describe several applications of Theorem \ref{thm:weighted-binary-join} 
to the weighted double-cone $\widetilde{K}_{2} + G$, for any regular graph $G$. For notational simplicity,
let $K_{2}^{b}$ denote $K_{2}$ if $b=1$ and $\overline{K}_{2}$ if $b=0$. \\

\par\noindent{\em Remark}: 
The next corollary complements the observation made by Casaccino \etal \cite{clms09} on $K_{2} + K_{m}$ 
where each vertex of $K_{2}$ has a weighted self-loop. They show that perfect state transfer occurs
in this weighted graph in contrast to the unweighted version.

\begin{corollary} \label{cor:weighted-double-cone}
For any $k$-regular graph $G$ on $n$ vertices and any $b \in \zo$, 
there exist weights $\mu,\eta \in \mathbb{R}^{+}$ so that the double-cone $\widetilde{K}^{b}_{2}(\mu,\eta)+G$ 
has perfect state transfer between the two vertices of $\widetilde{K}^{b}_{2}$.
\end{corollary}
\prf
Consider the weighted double-cone $\widetilde{K}^{b}_{2}(\mu,\eta) + \widetilde{G}(0,1)$,
where $\widetilde{G}(0,1)$ is simply the unweighted graph $G$.
We know that $\widetilde{K}^{b}_{2}(\mu,\eta)$ has perfect state transfer for 
$b\eta \tstar = (2\mathbb{Z}+1)\pi/2$. Note that when $b=0$, the perfect state transfer time is $\infty$
or non-existent.
Let $\delta = (\mu+b\eta)-k$ and $\Delta^{2} = \delta^{2} +8n$.
By Theorem \ref{thm:weighted-binary-join}, it suffices to have
\begin{equation}
\cos\left(\frac{\delta}{2}\tstar\right)\cos\left(\frac{\Delta}{2}\tstar\right) =
\cos\left(\frac{\delta}{4\eta}\pi\right)
\cos\left(\frac{\Delta}{4\eta}\pi\right) = (-1)^{1-b}.
\end{equation}
So, we require that:
\begin{enumerate}
\item $A \dotdef \delta/4\eta$ be an integer;
\item $B \dotdef \Delta/4\eta$ be an integer; and
\item $\iverson{A \equiv B\pmod{2}} = b$; or that $A$ and $B$ have the same parity if and only if $b=1$.
\end{enumerate}
From the first two conditions, we require that $\delta/\Delta$ be a rational number
$p/q < 1$ with $\gcd(p,q)=1$. 
Restating this last condition on $p$ and $q$ and simplifying, we get that
\begin{equation}
\delta \ = \ p \ \sqrt{\frac{8n}{q^{2}-p^{2}}}, \hspace{.5in}
\Delta \ = \ q \ \sqrt{\frac{8n}{q^{2}-p^{2}}}.
\end{equation}
So, we may choose
\begin{equation}
\eta \ = \ \frac{1}{4}\sqrt{\frac{8n}{q^{2}-p^{2}}}
\end{equation}
so that both $\delta/4\eta$ and $\Delta/4\eta$ are integers.
Therefore, we choose integers $p$ and $q$ satisfying $\gcd(p,q)=1$ and $\iverson{p \equiv q\pmod{2}} = b$;
this will satisfy all three conditions stated above.
Finally, we may choose $\mu = b\eta-k-\delta$ to complete the weight parameters.
\qed\\

\subsection{Double half-cones}

In this section, we consider graphs obtained by taking a half-join between $K_{2}$ and $G+G$, for some
arbitrary $k$-regular graph $G$, where each vertex of $K_{2}$ is connected to only one copy of $G$ in
the join $G+G$. When $G = \overline{K}_{n}$, this half-join is obtained by selecting two adjacent 
vertices in the complete bipartite graph $K_{n+1,n+1}$. 
In contrast to complete graphs, we show that weights are not helpful in complete bipartite graphs 
for achieving perfect state transfer. In fact, we prove a stronger result where perfect state
transfer still does not exist even if weights are added to some of the other sets of edges.

\begin{figure}[t]
\begin{center}
\setlength{\unitlength}{0.75cm}
\begin{picture}(10,5)
\thicklines
\put(2,1){\circle*{0.35}}
\put(2,3){\circle*{0.35}}
\put(2,1){\color{blue}{\line(0,1){2.0}}}
\put(1.65,1){\color{red}{\bigcircle{0.75}}}
\put(1.65,3){\color{red}{\bigcircle{0.75}}}
\put(0.75,0.9){\text{$\mu$}}
\put(0.75,2.9){\text{$\mu$}}
\put(1.5,1.9){\text{$\eta$}}
\put(4,1){\circle*{0.35}}
\put(4,3){\circle*{0.35}}
\put(6,1){\circle*{0.35}}
\put(6,3){\circle*{0.35}}
\put(8,1){\circle*{0.35}}
\put(8,3){\circle*{0.35}}
\put(2,1){\line(1,1){2.0}}
\put(2,3){\line(1,-1){2.0}}
\put(2,1){\line(2,1){4.0}}
\put(2,3){\line(2,-1){4.0}}
\put(2,1){\line(3,1){6.0}}
\put(2,3){\line(3,-1){6.0}}
\put(4,1){\line(0,1){2.0}}
\put(4,1){\line(1,1){2.0}}
\put(4,1){\line(2,1){4.0}}
\put(4,3){\line(1,-1){2.0}}
\put(4,3){\line(2,-1){4.0}}
\put(6,1){\line(0,1){2.0}}
\put(6,1){\line(1,1){2.0}}
\put(6,3){\line(1,-1){2.0}}
\put(8,1){\line(0,1){2.0}}
\end{picture}
\caption{Weighted half-join between $K_{2}$ and $K_{3,3}$. This is equivalent to adding weights
to a connected pair of vertices in the complete bipartite graph $K_{4,4}$.
There is no perfect state transfer between the two vertices with weighted self-loops. 
}
\end{center}
\end{figure}
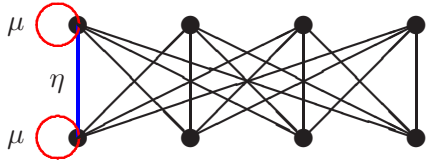

\begin{theorem} \label{thm:half-join}
Let $G$ be a $k$-regular graph on $n$ vertices. Let $\mathcal{G}(\mu,\eta;\kappa,\tau,\rho;\varepsilon)$ be a graph 
obtained from $\widetilde{K}_{2}(\mu,\eta)$ and $\widetilde{G}(\kappa,\tau) +_{\rho} \widetilde{G}(\kappa,\tau)$ 
by connecting each vertex of $\widetilde{K}_{2}(\mu,\eta)$ to exactly one copy of $\widetilde{G}(\kappa,\tau)$ 
in the weighted join $\widetilde{G}(\kappa,\tau) +_{\rho} \widetilde{G}(\kappa,\tau)$ 
and assigning a weight to $\varepsilon$ to each of these connecting edges.
Then, there are no non-zero real-valued weights $\mu$, $\eta$, $\kappa$, $\tau$, $\rho$ or $\varepsilon$ 
for which $\mathcal{G}(\mu,\eta;\kappa,\tau,\rho;\varepsilon)$ has perfect state transfer between the two 
vertices of $\widetilde{K}_{2}(\mu,\eta)$.
\end{theorem}

\par\noindent{\em Remark}: Note that if $\varepsilon = 0$, then we have perfect state transfer in
$\mathcal{G}$ trivially.\\

\prf
The adjacency matrix of $\mathcal{G}$ is given by
\begin{equation}
A_{\mathcal{G}} = 
\begin{bmatrix}
\mu  & \eta & \varepsilon \mathbf{1}_{n}^{T} & \mathbf{0}_{n}^{T} \\
\eta  & \mu & \mathbf{0}_{n}^{T} & \varepsilon \mathbf{1}_{n}^{T} \\
\varepsilon \mathbf{1}_{n} & \mathbf{0}_{n} & \kappa I_{n} + \tau A_{G} & \rho J_{n} \\
\mathbf{0}_{n} & \varepsilon \mathbf{1}_{n} & \rho J_{n} & \kappa I_{n} + \tau A_{G}
\end{bmatrix}
\end{equation}
where $\mathbf{0}_{n}$ and $\mathbf{1}_{n}$ denote the all-zero and all-one column vectors of dimension $n$, 
respectively.
Suppose that $A_{G}\ket{u_{j}} = \lambda_{j}\ket{u_{j}}$ are the eigenvalues and eigenvectors
of $G$, for $0 \le j \le n-1$, with $\ket{u_{0}}$ being the all-one eigenvector with $\lambda_{0} = k$.
Then, the spectra of $A_{\mathcal{G}}$ is given by the following sets:
\begin{enumerate}
\item The eigenvectors $\ket{0,0,0_{n},u_{j}}$ and $\ket{0,0,u_{j},0_{n}}$ both share the eigenvalues 
	$\kappa + \tau\lambda_{j}$, for $1 \le j \le n-1$. 
\item Let
	\begin{equation}
	\alpha_{\pm} = \frac{1}{2}(\delta_{\alpha} \pm \Delta_{\alpha}),
	\end{equation}
	where $\delta_{\alpha} = (\mu+\eta)-(\kappa + \tau k + \rho n)$ and 
	$\Delta_{\alpha}^{2} = \delta_{\alpha}^{2}+4\varepsilon^{2}n$.
	Then, the two eigenvectors 
	\begin{equation}
	\ket{\alpha_{\pm}} 
	= \frac{1}{\sqrt{L^{\alpha}_{\pm}}}
	\begin{bmatrix} \alpha_{\pm} & \alpha_{\pm} & 1_{n} & 1_{n} \end{bmatrix}^{T}
	\end{equation}
	have $\lambda_{\pm} = \alpha_{\pm}+(\kappa + \tau k + \rho n)$ as eigenvalues.
	Here $L^{\alpha}_{\pm} = 2(\alpha_{\pm})^{2}+2n$ is the normalization constant.
\item 
	Let
	\begin{equation}
	\beta_{\pm} = \frac{1}{2}(\delta_{\beta} \pm \Delta_{\beta}),
	\end{equation}
	where $\delta_{\beta} = (\mu-\eta)-(\kappa + \tau k -\rho n)$ and 
	$\Delta_{\beta}^{2} = \delta_{\beta}^{2}+4\varepsilon^{2}n$.
	Then, the two eigenvectors 
	\begin{equation}
	\ket{\beta_{\pm}} 
	= \frac{1}{\sqrt{L^{\beta}_{\pm}}}
	\begin{bmatrix} \beta_{\pm} & -\beta_{\pm} & 1_{n} & -1_{n} \end{bmatrix}
	\end{equation}
	have $\theta_{\pm} = \beta_{\pm}+(\kappa + \tau k - \rho n)$ as eigenvalues.
	Here $L^{\beta}_{\pm} = 2(\beta_{\pm})^{2}+2n$ is the normalization constant.
\end{enumerate}
The following identities can be verified easily: for $\idx \in \{\alpha,\beta\}$, we have
\begin{eqnarray}
L^{\idx}_{+}L^{\idx}_{-} & = & 4n\Delta^{2}_{\idx}/\varepsilon^{2} \\
\idx_{+}\idx_{-} & = & -n \\
\idx_{\pm}^{2}L^{\idx}_{\mp} & = & nL^{\idx}_{\pm} 
\end{eqnarray}
Using these, the quantum walk on $\mathcal{G}$ starting at $a$ and ending at $b$ is given by:
\begin{eqnarray}
\bra{b}e^{-itA_{\mathcal{G}}}\ket{a}
	& = & 
		\left\{\sum_{\pm} e^{-it\lambda_{\pm}}\frac{\alpha^{2}_{\pm}}{L^{\alpha}_{\pm}}\right\}
		-	
		\left\{\sum_{\pm} e^{-it\theta_{\pm}}\frac{\beta^{2}_{\pm}}{L^{\beta}_{\pm}}\right\}
\end{eqnarray}
After simplifications, we obtain
\begin{eqnarray}
\bra{b}e^{-itA_{\mathcal{G}}}\ket{a}
	& = &
	\frac{e^{-i(\kappa + \tau k)t}}{2} e^{-i(\rho n)t}e^{-i\delta_{\alpha}t/2} 
	\left[ \cos\left(\frac{\Delta_{\alpha}}{2}t\right) 
		- i\frac{\delta_{\alpha}}{\Delta_{\alpha}}\sin\left(\frac{\Delta_{\alpha}}{2}t\right) \right] \\
	& - &
	\frac{e^{-i(\kappa + \tau k)t}}{2} e^{i(\rho n)t}e^{-i\delta_{\beta}t/2}
	\left[ \cos\left(\frac{\Delta_{\beta}}{2}t\right) 
		- i\frac{\delta_{\beta}}{\Delta_{\beta}}\sin\left(\frac{\Delta_{\beta}}{2}t\right) \right]
\end{eqnarray}
Ignoring the irrelevant phase factor $e^{-i(\kappa + \tau k)t}$ and noting that the damping factor $\delta/\Delta$ 
forces the sine term to vanish, we get
\begin{eqnarray}
\bra{b}e^{-itA_{\mathcal{G}}}\ket{a}
	& = &
	\frac{e^{-i(\rho n)t}}{2} \cos\left(\frac{\delta_{\alpha}}{2}t\right) \cos\left(\frac{\Delta_{\alpha}}{2}t\right) 
	-
	\frac{e^{i(\rho n)t}}{2} \cos\left(\frac{\delta_{\beta}}{2}t\right) \cos\left(\frac{\Delta_{\beta}}{2}t\right) 
\end{eqnarray}
We choose $\tstar$ so that $e^{-i(\rho n)\tstar} = 1$, which implies that $\tstar = 2\mathbb{Z}\pi/\rho n$. 
This simplifies the above expression to
\begin{eqnarray}
\bra{b}e^{-i\tstar A_{\mathcal{G}}}\ket{a}
	& = &
	\frac{1}{2} \cos\left(\frac{\delta_{\alpha}}{2}\tstar\right) \cos\left(\frac{\Delta_{\alpha}}{2}\tstar\right) 
	-
	\frac{1}{2} \cos\left(\frac{\delta_{\beta}}{2}\tstar\right) \cos\left(\frac{\Delta_{\beta}}{2}\tstar\right) 
\end{eqnarray}
For simplicity, define
\begin{eqnarray}
Z_{\alpha} & = & \cos\left(\frac{\delta_{\alpha}}{2}\tstar\right) \cos\left(\frac{\Delta_{\alpha}}{2}\tstar\right) 
	= \cos\left(\frac{\delta_{\alpha}}{\rho n}\pi\right) \cos\left(\frac{\Delta_{\alpha}}{\rho n}\pi\right) \\
Z_{\beta}  & = & \cos\left(\frac{\delta_{\beta}}{2}\tstar\right) \cos\left(\frac{\Delta_{\beta}}{2}\tstar\right) 
	= \cos\left(\frac{\delta_{\beta}}{\rho n}\pi\right) \cos\left(\frac{\Delta_{\beta}}{\rho n}\pi\right)
\end{eqnarray}
Let 
\begin{equation}
\widetilde{P}_{\alpha} = \frac{\delta_{\alpha}}{\rho n}, \hspace{.25in}
\widetilde{Q}_{\alpha} = \frac{\Delta_{\alpha}}{\rho n}, \hspace{.25in}
\widetilde{P}_{\beta} = \frac{\delta_{\beta}}{\rho n}, \hspace{.25in}
\widetilde{Q}_{\beta} = \frac{\Delta_{\beta}}{\rho n}.
\end{equation}
To achieve perfect state transfer, we require that 
$Z_{\alpha}Z_{\beta} = -1$. For example, if we require $Z_{\alpha} = -1$ and $Z_{\beta} = 1$, 
then it suffices to impose the following {\em integrality} and {\em parity} conditions:
\begin{eqnarray} 
\widetilde{P}_{\alpha}, \widetilde{Q}_{\alpha} & \in & \mathbb{Z},
\hspace{.5in}
\widetilde{P}_{\alpha} \not\equiv \widetilde{Q}_{\alpha}\hspace{-0.1in}\pmod{2} \\
\widetilde{P}_{\beta}, \widetilde{Q}_{\beta} & \in & \mathbb{Z},
\hspace{.5in}
\widetilde{P}_{\beta} \equiv \widetilde{Q}_{\beta}\hspace{-0.1in}\pmod{2}.
\end{eqnarray}
We will show that there is no $\rho$ which can satisfy all the above conditions.

Suppose that, for $\idx \in \{\alpha, \beta\}$, we have 
\begin{equation}
\frac{\delta_{\idx}}{\Delta_{\idx}} \ = \ \frac{p_{\idx}}{q_{\idx}} \ \in \ \mathbb{Q}, 
\end{equation}
where $p_{\idx}$ and $q_{\idx}$ are integers with $\gcd(p_{\idx},q_{\idx})=1$; moreover, since
$\Delta_{\idx}^{2} = \delta_{\idx}^{2} + 4\varepsilon^{2}n$, we get
\begin{equation}
\delta_{\idx} = p_{\idx} \sqrt{\frac{4\varepsilon^{2}n}{q_{\idx}^{2} - p_{\idx}^{2}}},
\hspace{.5in}
\Delta_{\idx} = q_{\idx} \sqrt{\frac{4\varepsilon^{2}n}{q_{\idx}^{2} - p_{\idx}^{2}}}.
\end{equation}

\par\noindent
Consider $\widetilde{P}_{\idx}$ and $\widetilde{Q}_{\idx}$, for $\idx \in \{\alpha,\beta\}$.
Letting $\Lambda = 2\varepsilon/\rho\sqrt{n}$, we have
\begin{equation} \label{eqn:PQ}
\widetilde{P}_{\idx} = p_{\idx} \frac{\Lambda}{\sqrt{q_{\idx}^{2}-p_{\idx}^{2}}},
\hspace{.5in}
\widetilde{Q}_{\idx} = q_{\idx} \frac{\Lambda}{\sqrt{q_{\idx}^{2}-p_{\idx}^{2}}}. 
\end{equation}

\par
Since $\widetilde{P}_{\alpha}\equiv\widetilde{P}_{\alpha}^2\pmod{2}$, 
we know $\widetilde{P}_{\alpha}^{2}\not\equiv\widetilde{Q}_{\alpha}^{2}\pmod{2}$ is equivalent to 
$\widetilde{P}_{\alpha}\not\equiv\widetilde{Q}_{\alpha}\pmod{2}$.  Likewise, 
$\widetilde{P}_{\beta}^{2}\equiv\widetilde{Q}_{\beta}^{2}\pmod{2}$ is equivalent to 
$\widetilde{P}_{\beta}\equiv\widetilde{Q}_{\beta}\pmod{2}$.  
This changes (\ref{eqn:PQ}) to

\begin{equation} \label{eqn:PQ-squared}
\widetilde{P}_{\idx}^{2} = p_{\idx}^{2} \frac{\Lambda^{2}}{q_{\idx}^{2}-p_{\idx}^{2}},
\hspace{.5in}
\widetilde{Q}_{\idx}^{2} = q_{\idx}^{2} \frac{\Lambda^{2}}{q_{\idx}^{2}-p_{\idx}^{2}}. 
\end{equation}
Since we require that $\widetilde{P}_{\idx}$ and $\widetilde{Q}_{\idx}$ must be integers, 
then  $(q_{\idx}^{2}-p_{\idx}^{2}) \ | \ q_{\idx}^{2}\Lambda^{2}$ and 
$(q_{\idx}^{2}-p_{\idx}^{2}) \ | \ p_{\idx}^{2}\Lambda^{2}$.  However, $gcd(p_{\idx},q_{\idx})=1$ 
implies that $gcd(p_{\idx}^{2},q_{\idx}^{2})=1$.  
This gives us $(q_{\idx}^{2}-p_{\idx}^{2}) \ | \ \Lambda^{2}$.

Suppose now that $p_{\beta}^{2}\equiv{q_{\beta}^{2}}\pmod{2}$.  
Then $q_{\beta}^{2}-p_{\beta}^{2}$ is even.  
This forces $\Lambda^{2}$ to be even.  
Similarly, suppose $p_{\beta}^{2}\not\equiv{q_{\beta}^{2}}\pmod{2}$.
Then $q_{\beta}^{2}-p_{\beta}^{2}$ is odd.  
However, since $\widetilde{P}_{\beta}^{2}\equiv\widetilde{Q}_{\beta}^{2}\pmod{2}$
and one of $p_{\beta}^{2}$,$q_{\beta}^{2}$ is odd, then $\Lambda^{2}$ must be even.

In both cases, $\Lambda^{2}$ is even.  
Allowing $p_{\alpha}^{2}\equiv{q_{\alpha}^{2}}\pmod{2}$ guarantees 
$\widetilde{P}_{\alpha}^{2}\equiv\widetilde{Q}_{\alpha}^{2}\pmod{2}$.  
Letting $p_{\alpha}^{2}\not\equiv{q_{\alpha}^{2}}\pmod{2}$ gives us 
$q_{\alpha}^{2}-p_{\alpha}^{2}$ to be odd.  
This again forces $\widetilde{P}_{\alpha}^{2}\equiv\widetilde{Q}_{\alpha}^{2}\pmod{2}$.  
Both instances contradict our given requirement that 
$\widetilde{P}_{\alpha}^{2}\not\equiv\widetilde{Q}_{\alpha}^{2}\pmod{2}$.

The case when we require that $Z_{\alpha} = 1$ and $Z_{\beta} = -1$, that is, where
$\widetilde{P}_{\alpha}$ is even and $\widetilde{Q}_{\alpha}$, $\widetilde{P}_{\beta}$, $\widetilde{Q}_{\beta}$ 
are odd, may be treated similarly.
\qed\\

\begin{corollary}
For any $n \ge 2$, consider the complete bipartite graph $K_{n,n}$. Let $a$ and $b$ be two arbitrary
adjacent vertices in $K_{n,n}$. Then, there are no self-loop weights $\mu$ on $a$ and $b$ and edge weight 
$\eta$ on the edge $(a,b)$ for which there is perfect state transfer from vertex $a$ to vertex $b$ in this
weighted version of $K_{n,n}$.
\end{corollary}
\prf
We apply Theorem \ref{thm:half-join} with $G=\overline{K}_{n-1}$ set to the empty graph on $n-1$ vertices, 
that is $A_{G}$ is the all-zero matrix and hence $k = 0$. Also, we set $\varepsilon = 1$, $\kappa = 0$ and 
$\tau$ is an arbitrary value.
In the proof of Theorem \ref{thm:half-join}, setting $\kappa = 0$ does not affect perfect state transfer 
since the term $\kappa + k\tau$ may be ignored due to its contribution as a global phase factor.
Setting $\varepsilon = 1$ does not affect perfect state transfer since it is ``factored out'' through $\Lambda$.
Thus, these specific setting of values do not affect the conclusions of Theorem \ref{thm:half-join}.
\qed


\section{Hamming graphs}

We show that the class of weighted Hamming graphs exhibit perfect state transfer between any two of its vertices.
First, we prove the following closure result on Cartesian product of graphs. This is an adaptation of a similar
theorem for the unweighted case (see \cite{anoprt09}).

\begin{theorem} \label{thm:weighted-cartesian}
For $j=1,\ldots,m$, the graph $G_{j}$ has perfect state transfer from $a_{j}$ to $b_{j}$ at time $t_{j}$ 
if and only if $\mathcal{G} = \bigoplus_{j=1}^{m} \widetilde{G}_{j}(\mu_{j},\eta_{j})$ has perfect state transfer 
from $(a_{1},\ldots,a_{m})$ to $(b_{1},\ldots,b_{m})$ at time $\tstar$, 
whenever $\tstar = t_{j}/\eta_{j}$. This holds independently of the choice of the self-loop weights $\mu_{j}$.
\end{theorem}
\prf
We prove the claim for $m=2$. 
Suppose that the unweighted graph $G_{j}$ has perfect state transfer from $a_{j}$ to $b_{j}$ at time $\tstar_{j}$.
Consider the quantum walk on the $\widetilde{G}_{1}(\mu_{1},\eta_{1}) \oplus \widetilde{G}_{2}(\mu_{2},\eta_{2})$.
For shorthand, we denote each graph simply as $\widetilde{G}_{j}$:
\begin{eqnarray}
\bra{b_{1},b_{2}} e^{-itA_{\widetilde{G}_{1} \oplus \widetilde{G}_{2}}}\ket{a_{1},a_{2}}
	& = & \bra{b_{1}}\bra{b_{2}}
			e^{-it(I \otimes A_{\widetilde{G}_{2}})}e^{-it(A_{\widetilde{G}_{1}} \otimes I)}
			\ket{a_{1}}\ket{a_{2}} \\
	& = & \bra{b_{1}}\bra{b_{2}}
			(I \otimes e^{-it A_{\widetilde{G}_{2}}})
			(e^{-it A_{\widetilde{G}_{1}}} \otimes I)
			\ket{a_{1}}\ket{a_{2}} \\
	& = & \bra{b_{1}}e^{-it A_{\widetilde{G}_{1}}}\ket{a_{1}}
			\bra{b_{2}}e^{-it A_{\widetilde{G}_{2}}}\ket{a_{2}}.
\end{eqnarray}
Since $A_{\widetilde{G}(\mu,\eta)} = \mu I + \eta A_{G}$, we have
\begin{equation}
\bra{b}e^{-it A_{\widetilde{G}}}\ket{a} = e^{-i\mu t}\bra{b}e^{-i\eta t A_{G}}\ket{a}.
\end{equation}
Therefore, the quantum walk on the weighted Cartesian product yields
\begin{eqnarray}
\bra{b_{1},b_{2}} e^{-itA_{\widetilde{G}_{1} \oplus \widetilde{G}_{2}}}\ket{a_{1},a_{2}}
	& = & e^{-i(\mu_{1}+\mu_{2})t}\bra{b_{1}}e^{-i\eta_{1}t A_{G_{1}}}\ket{a_{1}}
			\bra{b_{2}}e^{-i\eta_{2}t A_{G_{2}}}\ket{a_{2}}.
\end{eqnarray}
This shows that $\widetilde{G}_{1} \oplus \widetilde{G}_{2}$ has perfect state transfer from 
$(a_{1},a_{2})$ to $(b_{1},b_{2})$ at time $t$ if and only if 
$G_{1}$ has perfect state transfer from $a_{1}$ to $b_{1}$ at time $\eta_{1}t$
{\em and}
$G_{2}$ has perfect state transfer from $a_{2}$ to $b_{2}$ at time $\eta_{2}t$.
So, if the weights $\eta_{j}$ satisfy $\eta_{j}\tstar = t_{j}$, for all $j$, 
then $\widetilde{G}_{1} \oplus \widetilde{G}_{2}$ has perfect state transfer at time $\tstar$. 
The general claim follows by induction.
\qed\\

\begin{figure}[t]
\begin{center}
\setlength{\unitlength}{0.75cm}
\begin{picture}(15,5)
\thicklines

\put(2,1){\circle*{0.35}}
\put(2,5){\circle*{0.35}}
\put(6,1){\circle*{0.35}}
\put(6,5){\circle*{0.35}}
\put(2,1){\line(0,1){4.0}}
\put(2,1){\line(1,0){4.0}}
\put(6,5){\line(-1,0){4.0}}
\put(6,5){\line(0,-1){4.0}}
\put(3,2){\circle*{0.35}}
\put(3,4){\circle*{0.35}}
\put(5,2){\circle*{0.35}}
\put(5,4){\circle*{0.35}}
\put(3,2){\line(0,1){2.0}}
\put(3,2){\line(1,0){2.0}}
\put(5,4){\line(-1,0){2.0}}
\put(5,4){\line(0,-1){2.0}}
\put(2,1){\line(1,1){1.0}}
\put(2,5){\line(1,-1){1.0}}
\put(6,1){\line(-1,1){1.0}}
\put(6,5){\line(-1,-1){1.0}}

\put(9,1){\circle*{0.35}}
\put(11,1){\circle*{0.35}}
\put(13,1){\circle*{0.35}}
\put(9,3){\circle*{0.35}}
\put(11,3){\circle*{0.35}}
\put(13,3){\circle*{0.35}}
\put(9,5){\circle*{0.35}}
\put(11,5){\circle*{0.35}}
\put(13,5){\circle*{0.35}}
\put(9,1){\line(1,0){4.0}}
\put(9,1){\line(0,1){4.0}}
\put(13,5){\line(-1,0){4.0}}
\put(13,5){\line(0,-1){4.0}}
\put(11,1){\line(0,1){4.0}}
\put(9,3){\line(1,0){4.0}}
\curve(9,1,11,1.5,13,1)
\curve(9,3,11,3.5,13,3)
\curve(9,5,11,5.5,13,5)
\curve(9,1,9.5,3,9,5)
\curve(11,1,11.5,3,11,5)
\curve(13,1,13.5,3,13,5)

\end{picture}
\caption{Hamming graphs: (a) $H(2,3)$ (b) $H(3,2)$.
Perfect state transfer occurs between any pair of vertices with the help of weighted self-loops and edges.}
\end{center}
\end{figure}
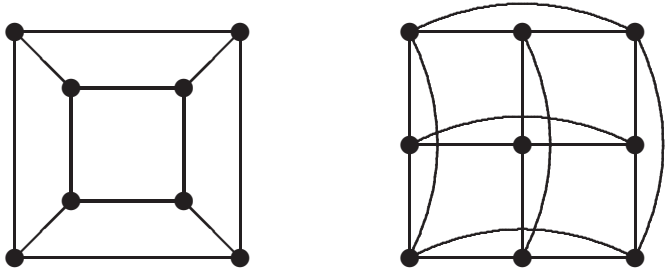

\begin{theorem} \label{thm:weighted-hamming}
The class $\widetilde{H}(q,n)$ of weighted Hamming graphs has universal perfect state transfer 
at an arbitrarily chosen time.
\end{theorem}
\prf
Recall that $H(q,n) = K_{q}^{\oplus n}$. 
Let $a=(a_{1},\ldots,a_{n})$ and $b=(b_{1},\ldots,b_{n})$ be two vertices of $\widetilde{H}(q,n)$. 
By Corollary \ref{cor:weighted-double-cone}, we know that $\widetilde{K}_{q}$ has perfect state transfer 
between any two of its vertices for a suitable choice of weights. 
For each dimension $j \in \{1,\ldots,n\}$, fix a set of weights so that $\widetilde{K}_{q}^{(j)}$ 
has perfect state transfer from $a_{j}$ to $b_{j}$. 
Then, by Theorem \ref{thm:weighted-cartesian}, $\bigoplus_{j=1}^{n} \widetilde{K}_{q}^{(j)}$ 
has perfect state transfer from $a$ to $b$.
\qed\\


\subsection{Hypercubes}

In this section, we show that a weighted hypercube has universal perfect state transfer property. 
In fact, we prove a stronger statement as given in the next theorem. But first, we need to define a 
particular notion of uniform superposition over the $n$-cube.

\begin{fact} \label{fact:hypercube} {\em (Moore-Russell \cite{mr02}, Bernasconi \etal \cite{bgs08})}\\
The following facts are known about a quantum walk on the hypercube $Q_{n}$ at times
$t \in \{\pi/4,\pi/2\}$:
\begin{equation}
\bra{b}e^{-it{Q_{n}}}\ket{a} = 
\left\{\begin{array}{ll}
	(-i)^{|a \oplus b|}/\sqrt{2^{n}} 	& \mbox{ if $t = \pi/4$ } \\
	\iverson{a \oplus b = 1_{n}} 				& \mbox{ if $t = \pi/2$ } 
	\end{array}\right.
\end{equation}
\end{fact}

\par\noindent We say that a superposition $\ket{\varrho_{n}}$ over $Q_{n}$ is in {\em normal form} if 
\begin{equation}
\ket{\varrho_{n}} = \frac{1}{\sqrt{2^{n}}} \sum_{a \in \zo^{n}} (-i)^{|a|}\ket{a}.
\end{equation}
Note that $\ket{\varrho_{n}}$ is the uniform superposition of a quantum walk on $Q_{n}$ from $0_{n}$
at time $\pi/4$; that is, $\ket{\varrho_{n}} = \exp(-i(\pi/4) Q_{n})\ket{0_{n}}$.

\begin{theorem} \label{thm:weighted-cube}
For any $n \ge 1$, given any two distinct subcubes $B_{1}$ and $B_{2}$ of $Q_{n}$, 
there is a set of edge weights $w$ so that $Q_{n}^{w}$ has perfect state transfer 
between uniform superpositions in normal form on $B_{1}$ and $B_{2}$.
\end{theorem}
\prf
First, we show that the hypercube $Q_{n}$ has perfect state transfer from any vertex to any subcube.
Since $Q_{n}$ is vertex-transitive, it suffices to show perfect state transfer from vertex $0_{n}$ to
the subcube $B = (1_{k}0_{\ell}\star_{m})$, where $m = n-k-\ell$. 
Define the adjacency matrix of $\widetilde{Q}_{n}$ as
\begin{equation}
\widetilde{Q}_{n} = Q_{k} \otimes I_{2^{n-k}} + \mbox{\small $\frac{1}{2}$} I_{2^{k+\ell}} \otimes Q_{m}.
\end{equation}
which is a sum of two commuting matrices. Then, letting $\tstar = \pi/2$, we have
\begin{equation}
\bra{1_{k}0_{\ell}}\bra{\varrho_{m}}\exp\left(-i\tstar\widetilde{Q}_{n}\right)\ket{0_{k}0_{\ell}0_{m}}
	= 
	\bra{1_{k}0_{\ell}}\bra{\varrho_{m}}
	\exp\left(-i\frac{\tstar}{2} I_{2^{k+\ell}} \otimes Q_{m}\right)
	\ket{1_{k}0_{\ell}0_{m}}.
\end{equation}
The equality and the fact that the last expression has unit magnitude follows from Fact \ref{fact:hypercube}.

To show perfect state transfer between two arbitrary subcubes, note that we just showed that
$\ket{B} = e^{-i\tstar\widetilde{Q}_{n}}\ket{0_{n}}$. 
Thus, we also have $\ket{0_{n}} = e^{-i\tstar(-\widetilde{Q}_{n})}\ket{B}$. This proves the claim.
\qed\\

\par\noindent
We recover the result of Bernasconi \etal \cite{bgs08}, which we restate in the next corollary, via the
use of explicit edge weights on the hypercube.

\begin{corollary}
For any $n \ge 1$, given any two distinct vertices $a$ and $b$ of the hypercube $Q_{n}$, there is a set of
edge weights $w$ so that $Q_{n}^{w}$ has perfect state transfer from $a$ to $b$ at time $\tstar = \pi/2$.
\end{corollary}

\par\noindent{\em Remark}:
We note that Bernasconi \etal \cite{bgs08} proved universal perfect state transfer for the $n$-cube by 
{\em dynamically} changing the underlying hypercubic structure of the graph. In contrast, our scheme is based
on using {\em static} weights which can be interpreted dynamically with time. In both schemes, it is possible
to route information through a Hamiltonian path which visits each vertex once and exactly once. 
We believe that this Hamiltonian property might be of interest in further applications of perfect state transfer.


\section{Conclusion}

We studied perfect state transfer on quantum networks represented by weighted graphs.
Our goal was to understand the role of weights in achieving perfect state transfer in graphs. 

First, we proved a join theorem for weighted regular graphs and derived, as a corollary, that a 
weighted double-cone on any regular graph has perfect state transfer. 
This implies as a corollary a result of Casaccino \etal \cite{clms09} where the regular graph is a complete graph.
In contrast, we also showed that weights do not help in achieving perfect state transfer in complete 
bipartite graphs. This is obtained as part of a more general result on graphs constructed from a half-join 
of $K_{2}$ and $G+G$, for an arbitrary regular graph $G$. We found it curious that the full join connection 
seemed crucial for weights to have a positive effect in achieving perfect state transfer.
We leave the case of complete multipartite graphs and strongly regular graphs as an open question.

Second, we observed that Hamming graphs have the universal perfect state transfer property. 
This is a stronger requirement that the standard perfect state transfer property where perfect state transfer
must occur between any pair of vertices. Prior to this work, the only known family of graphs with universal 
perfect state transfer were the (unweighted) hypercubic graphs \cite{bgs08}. 
We proved our result on the Hamming graphs by showing a closure result for a weighted Cartesian product of 
perfect state tranfer graphs; even when the graph components have different perfect state transfer times. 
The unweighted version of this closure result, as shown in \cite{anoprt09}, requires a global common 
perfect state transfer time for all graphs in the Cartesian product. 
For the hypercubes, we showed a stronger universal perfect state transfer property, where perfect state 
transfer occurs between uniform superpositions of two arbitrary subcubes. We imposed a mild condition on 
the uniform superpositions which exhibit perfect state transfer.

We remark that if zero weights are allowed, then universal perfect state transfer is trivial.
Simply take any path connecting the two vertices and assign the hypercubic weights to the edges
on the path (as in Christandl \etal \cite{cdel04}) and zero weights to all other edges. If zero
weights are not allowed then we conjecture that near-perfect state transfer is possible by
assigning weights that tend to zero (for the edges which require zero weights).

\begin{figure}[t]
\begin{center}
\setlength{\unitlength}{0.75cm}
\begin{picture}(17,5)
\thicklines
\put(3,1){\circle*{0.35}}
\put(1,2){\circle*{0.35}}
\put(3,2){\circle*{0.35}}
\put(5,2){\circle*{0.35}}
\put(1,3){\circle*{0.35}}
\put(3,3){\circle*{0.35}}
\put(5,3){\circle*{0.35}}
\put(3,4){\circle*{0.35}}
\put(3,1){\line(0,1){1.0}}
\put(3,1){\line(-2,1){2.0}}
\put(3,1){\line(2,1){2.0}}
\put(3,2){\line(-2,1){2.0}}
\put(3,2){\line(2,1){2.0}}
\put(3,4){\line(0,-1){1.0}}
\put(3,4){\line(-2,-1){2.0}}
\put(3,4){\line(2,-1){2.0}}
\put(3,3){\line(-2,-1){2.0}}
\put(3,3){\line(2,-1){2.0}}
\put(1,2){\line(0,1){1.0}}
\put(5,2){\line(0,1){1.0}}

\put(9.5,1){\circle*{0.35}}
\put(9.5,2){\circle*{0.35}}
\put(9.5,3){\circle*{0.35}}
\put(9.5,4){\circle*{0.35}}
\put(9.5,1){\line(0,1){3.0}}
\put(9.5,2){\line(0,1){1.0}}
\put(8,1.25){\text{\small $\sqrt{1 \cdot 3}$}}
\put(8,2.25){\text{\small $\sqrt{2 \cdot 2}$}}
\put(8,3.25){\text{\small $\sqrt{3 \cdot 1}$}}

\matrixput(13,1)(1,0){4}(0,1){4}{\circle*{0.35}}
\linethickness{0.1pt}
\matrixput(13,1)(1,0){3}(0,1){4}{\line(1,0){1.0}}
\matrixput(13,1)(1,0){4}(0,1){3}{\line(0,1){1.0}}
\linethickness{1.5pt}
\put(13,1){\color{blue}{\line(1,0){2.0}}}
\put(15,1){\color{blue}{\line(0,1){3.0}}}
\put(15,4){\color{blue}{\line(1,0){1.0}}}
\put(13.05,0.25){\text{\scriptsize $\sqrt{6}$}}
\put(14.05,0.25){\text{\scriptsize $\sqrt{10}$}}
\put(15.05,1.25){\text{\scriptsize $\sqrt{12}$}}
\put(15.05,2.25){\text{\scriptsize $\sqrt{12}$}}
\put(15.05,3.25){\text{\scriptsize $\sqrt{10}$}}
\put(15.05,4.25){\text{\scriptsize $\sqrt{6}$}}

\end{picture}
\caption{Existence of universal near-perfect state transfer on any weighted graph.
(a) $Q_{n}$ has vertex-to-vertex PST (Bernasconi \etal \cite{bgs08}) 
(b) Hypercubic-weighted $P_{n}$ has end-to-end PST (Christandl \etal  \cite{cdel04}) 
(c) Emulate the hypercubic weighting along any path between the source and target vertices
while setting other weights to near zero.
}
\end{center}
\end{figure}
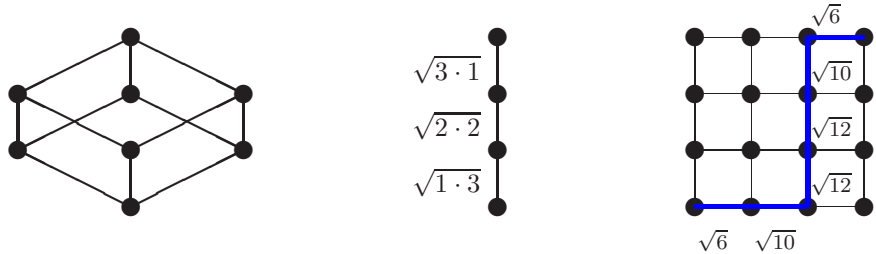


\section*{Acknowledgments}

This research was supported in part by the National Science Foundation grant DMS-0646847
and also by the National Security Agency grant H98230-09-1-0098.


\end{document}